\documentclass[ twocolumn,
aps,prd,   
               preprintnumbers,numbers,sort&compress,nofootinbib,
                            showpacs,
               colorlinks,
               linkcolor=blue,
               citecolor=blue]{revtex4-1}    
         \newcommand{\exclude}[1]{}

\usepackage{graphicx,amsmath,amssymb,bm}
\usepackage{psfrag}
\usepackage{hyperref}
\newcommand{\be}{\begin{eqnarray}}
\newcommand{\ee}{\end{eqnarray}}

\begin{document}
\title{Quark (Anti)Nugget Dark Matter}

\author{Kyle Lawson and Ariel R. Zhitnitsky}
 \affiliation{Department of Physics \& Astronomy, University of British Columbia, Vancouver, B.C. V6T 1Z1, Canada}

\begin{abstract}
We review a  testable 
dark matter model outside of the standard WIMP paradigm in which the 
observed ratio $\Omega_{\rm dark} \simeq 5\cdot \Omega_{\rm visible}$ for 
visible and dark matter 
densities  finds its natural explanation  as a result
of their common QCD origin. 
Special emphasis is placed on the observational consequences 
of this model and on the detection prospects for 
present or planned experiments. In particular, we argue that the relative intensities for a number of observed 
excesses of emission  (covering more than 11 orders of magnitude) can be explained by this model without any new fundamental parameters 
as all relative intensities for these   emissions are determined by  standard and well established physics.    
 
 \end{abstract}
 \maketitle

\section{QCD as a single source for Dark Matter and visible Baryons}\label{DM}
In this proposal we argue that two of the largest open questions in 
cosmology, the origin of the matter/antimatter asymmetry and the 
nature of the dark matter (DM), may have their origin within a single 
theoretical framework. Furthermore, both effects  may originate at the same 
cosmological epoch  from one and the same QCD physics. 

It is generally assumed that the universe 
began in a symmetric state with zero global baryonic charge 
and later, through some baryon number violating process, 
evolved into a state with a net positive baryon number. As an 
alternative to this scenario we advocate a model in which 
``baryogenesis'' is actually a charge separation process 
in which the global baryon number of the universe remains 
zero. In this model the unobserved antibaryons come to comprise 
the dark matter.  A connection between dark matter and 
baryogenesis is made particularly compelling by the 
similar energy densities of the visible and dark matter 
with $\Omega_{\rm dark} \simeq 5\cdot \Omega_{\rm visible}$. If these processes 
are not fundamentally related the two components could 
exist at vastly different scales. 

\begin{figure}[t]
\begin{center}
\includegraphics[width = 0.5\textwidth]{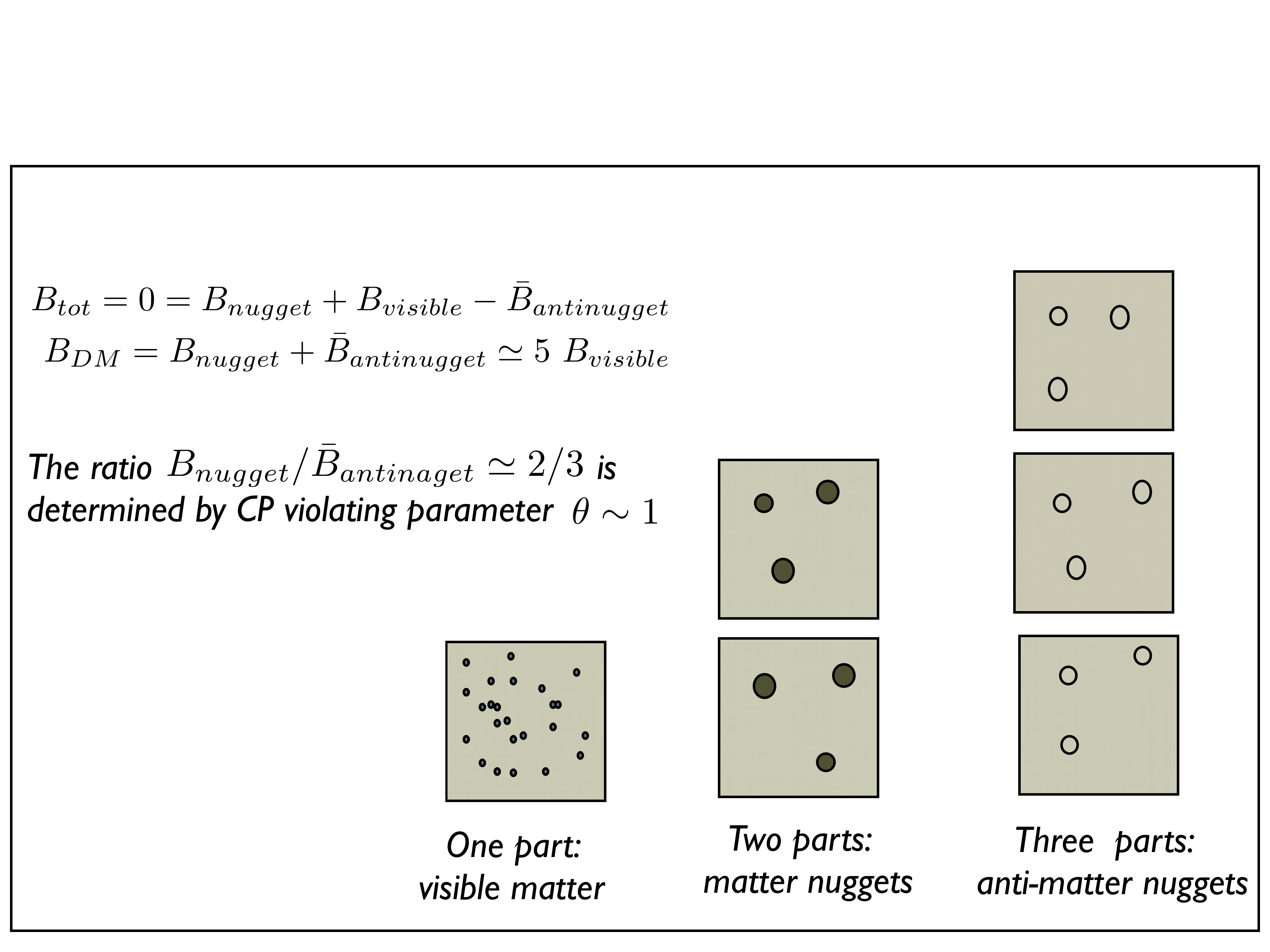}
\caption{ Matter in the Universe. A model which explains both the 
matter -antimatter asymmetry and the observed ratio of visible matter and DM}
\label{fig:matter}
\end{center}
\end{figure}

In this model baryogenesis occurs at 
the QCD phase transition. Both quarks and antiquarks are 
thermally abundant in the primordial plasma but, in 
addition to forming conventional baryons, some fraction 
of them are bound into heavy nuggets of quark matter in a 
colour superconducting phase. Nuggets of both matter and 
antimatter are formed as a result of the dynamics of the axion domain walls \cite{Zhitnitsky:2002qa,Oaknin:2003uv}, some details of this process will be discussed  
in section \ref{nuggets}. Were CP symmetry to be exactly preserved  
an equal number of matter and antimatter nuggets would form resulting in 
no net ``baryogenesis". However, CP violating processes associated 
with the axion $\theta$ term in QCD result in the preferential formation of 
antinuggets 
\footnote{This preference  is essentially determined by the 
sign of $\theta$. Note, that  the  idea of a charge separation mechanism 
resulting from local violation of  
\textsc{CP} invariance through an induced $\theta_{ind}$ can be 
experimentally tested  at the Relativistic Heavy Ion Collider (RHIC) and 
the LHC. We include a few comments  and relevant references, including 
some references to recent experimental results supporting the basic idea,  in section \ref{conclusion}.}.  
At the phase transition $\theta \sim 1$ and all asymmetry effects would 
have been order one while during the
present epoch, long after the phase transition, this 
source of \textsc{CP} violation is no longer available. The remaining 
antibaryons in the plasma then annihilate away leaving only the baryons 
whose antimatter counterparts are bound in the excess of antinuggets and thus 
unavailable to annihilate. The observed 
matter to dark matter ratio results if the  number  of 
antinuggets is larger than number of nuggets 
by a factor of $\sim$ 3/2 at the end of nugget formation. This would 
result in a matter content with baryons, quark nuggets 
and antiquark nuggets in an approximate  ratio 
\be
\label{ratio}
B_{\rm visible}: B_{\rm nuggets}: B_{\rm antinuggets}\simeq 1:2:3, 
\ee
and  no net baryonic charge, as sketched   on Fig.\ref{fig:matter}.

Unlike conventional dark matter candidates, dark-matter/antimatter
nuggets are strongly interacting but macroscopically large.  
They do not contradict the many known observational
constraints on dark matter or
antimatter  for three main reasons~\cite{Zhitnitsky:2006vt}:
\begin{itemize} 
\item They carry a huge (anti)baryon charge 
$|B|  \gtrsim 10^{25}$, and so have an extremely tiny number
density; 
\item The nuggets have nuclear densities, so their effective interaction
is small $\sigma/M \sim 10^{-10}$ ~cm$^2$/g,  well below the typical astrophysical
and cosmological limits which are on the order of 
$\sigma/M<1$~cm$^2$/g;
\item They have a large binding energy 
such that the baryon charge  in the
nuggets is not available to participate in big bang nucleosynthesis
(\textsc{bbn}) at $T \approx 1$~MeV. 
\end{itemize} 
To reiterate: the weakness of the visible-dark matter interaction 
in this model due to the small geometrical parameter $\sigma/M \sim B^{-1/3}$ 
rather than due to the weak coupling 
of a new fundamental field to standard model particles. 
It is this small effective interaction $\sim \sigma/M \sim B^{-1/3}$ which replaces 
the conventional requirement of sufficiently weak interactions for WIMPs.

An fundamental measure of the scale of baryogenesis is the 
baryon to entropy ratio at the present time
\be
\label{eta}
\eta\equiv\frac{n_B-n_{\bar{B}}}{n_{\gamma}}\simeq \frac{n_B}{n_{\gamma}}\sim 10^{-10}.
\ee
If the nuggets were not present after the phase transition the conventional baryons 
and anti-baryons would continue to annihilate each other until the temperature 
reaches $T\simeq 22$ MeV when density would be 9 orders of magnitude smaller 
than observed. This annihilation catastrophe, normally thought to be  resolved as a result of  ``baryogenesis," is avoided in our proposal because  more  anti-baryons than 
baryons are hidden in the form of the macroscopical nuggets and thus no longer available 
for annihilation. Only the visible baryons (not anti-baryons)  remain in the system 
after nugget formation is fully completed.

In our proposal (in contrast with conventional models) the ratio $\eta$ is determined 
by the formation temperature $T_{\rm form}$ at 
which the nuggets and anti-nuggets basically 
have competed  their formation and below which annihilation with 
surrounding matter becomes negligible.    
This temperature is determined by many factors: transmission/reflection 
coefficients, evolution of the nuggets, expansion of the universe, cooling rates, evaporation 
rates, the  dynamics of the axion domain wall network, etc. 
In general, all of these effects will contribute contribute equally to 
determining $T_{\rm form}$ at  the QCD scale. Technically, the corresponding 
effects are hard    to compute as even basic properties of the  QCD phase diagram at nonzero 
$\theta$ are still unknown. However, an approximate estimate of $T_{\rm form}$  is 
quite simple as it must be  expressed in terms of the gap 
$\Delta\sim 100$ MeV when the colour 
superconducting  phase  sets in  inside the nuggets. The observed ratio (\ref{eta}) 
corresponds to $T_{\rm form}\simeq 41$ MeV 
which is  indeed a typical QCD scale slightly below the critical 
temperature $T_c\simeq 0.6 \Delta$ when colour superconductivity sets in. 

In different words, in this proposal the ratio (\ref{eta}) emerges as a result of the 
QCD dynamics when process of charge separation stops at  $T_{\rm form}\simeq 41$ 
MeV, rather than a result of baryogenesis when a net baryonic charge is produced.

\section{ Quark (anti)nuggets as Dark Matter}\label{nuggets}
The majority of dark matter models assume the existence of 
a new fundamental field coupled only weakly to the 
standard model particles, these models may then be tuned 
to match the observed dark matter properties. We take a different 
perspective and consider the possibility that the dark matter 
is in fact composed of well known quarks and antiquarks but in 
a new high density phase, similar to the Witten's strangelets~\cite{Witten:1984rs}.
The only new crucial element in comparison with previous studies based on 
Witten's droplets  ~\cite{Witten:1984rs}  is that the nuggets could be made of 
matter as well as antimatter in our framework, and 
the stability of the DM nuggets is provided by the axion domain 
walls \cite{Zhitnitsky:2002qa}. 

Though the QCD phase diagram at $\theta\neq 0$ 
is not known, it is well understood that $\theta$ is in fact the angular variable, 
and therefore supports various types of the domain walls, including the so-called 
$N=1$ domain walls when $\theta$ interpolates between one and the same 
physical vacuum state $\theta\rightarrow\theta+2\pi$. While such domain walls 
are formally unstable, their life time could be much longer than life time of the universe. 
Furthermore, it is expected that closed bubbles made of these $N=1$ axion domain 
walls are also produced during the QCD phase transition
with a typical correlation length $\sim m_{a}^{-1}$ where $m_a $ is the axion mass. 
The $N=1$ axion domain walls are unique in a sense that they might be formed even
in case of inflation which normally prevents the generation of any other types of topological defects.  

The collapse of these closed bubbles is halted due to the fermi pressure acting  
inside of the bubbles as sketched  on Fig. \ref{fig:nugget}.
The equilibrium of the obtained system has been analyzed in 
\cite{Zhitnitsky:2002qa} for a specific axion domain wall with  tension 
$\sigma_a\simeq 1.8\cdot 10^8 \rm{GeV}^3$ which corresponds to 
$m_a\sim 10^{-6}$ eV. For these  axion parameters it has been found that a 
typical baryon charge of the nugget is $B\sim 10^{32}$ while a typical 
size of the nugget is $R\sim 10^{-3}{\rm cm}$.
 Using the dimensional arguments one can easily infer that these    
 parameters   scale  with the axion mass as follows
\be
\sigma_a\sim m_a^{-1}, ~~~R\sim \sigma_a, ~~~~B\sim \sigma_a^3.
\ee
Therefore,  when   the axion mass 
$m_a$ varies within the observationally allowed window $10^{-6} {\rm eV}\leq m_a\leq  10^{-3}{\rm eV}$, see e.g. reviews  \cite{Asztalos:2006kz,Sikivie:2009fv}, the corresponding nuggetÕs parameters also vary as follows
\be
\label{B}
 10^{-6} {\rm cm}\lesssim R \lesssim 10^{-3}{\rm cm}, ~~~ 10^{23}\lesssim B\lesssim 10^{32}.
\ee
The corresponding allowed  region is essentially  uncovered by present  experiments, see Fig.\ref{fig:limits} from section \ref{direct}.

While the observable consequences of this model are strongly suppressed  
by the low number density of the quark nuggets the interaction of these objects 
with the visible matter of the galaxy will necessarily produce observable 
effects. Any such consequences will be largest where the densities 
of both visible and dark matter are largest such as in the 
core of galaxies or the early universe. The nuggets essentially 
behave as  conventional cold DM in an environment where the surrounding 
density is small, while in environments of sufficiently large density
they begin interacting and emitting radiation  
(i.e. effectively become visible matter)  when they are placed in an 
environment of sufficiently large density 
\footnote{In this short review  we concentrate on the phenomenological 
consequences of anti-nuggets which can act as additional radiation sources
as a result of rare annihilation events, see sections \ref{observations},
 \ref{direct} for some details.  Matter nuggets may also be phenomenologically  
interesting as they will occasionally be captured by astronomical objects. 
Captured matter nuggets will behave very differently 
from  strangelets~\cite{Witten:1984rs}  because  the nuggets do not  convert 
the entire surrounding object into strange matter. Rather, the high density regions 
inside any object will remain a finite size (\ref{B}) with the quark matter    
extending to eventually become normal nuclear matter at lower 
densities, as in studies of possible colour superconducting cores of 
neutron stars. The possibility that standard astronomical objects may have a small 
quark matter core implies that some objects may bebe observer to have a much 
higher density than typically assumed, see e.g. \cite{Labun:2011wn} and the many 
references therein on some observational consequences  of  high 
density objects dressed by a normal matter.}.
 
We emphasize that the phenomenologically relevant features of the nuggets 
are determined by properties of the surface layer of electrons (or 
positrons in the case of an antimatter nugget) known as the
electrosphere as sketched on lower right corner of Fig.\ref{fig:nugget}.
These properties are in principle, calculable from first principles using only 
the well established rules of QCD and QED. As such 
the model contains no tunable fundamental parameters, except for a single mean baryon number  $<B>$
which is hard to compute theoretically as it depends on all complications mentioned above such as QCD phase diagram at $\theta\neq 0$, formation and evolution of the nuggets, etc. This parameter $<B>\sim 10^{25}$  is fixed in our  proposal assuming that anti-nuggets saturate the observed  511 keV line from the centre of galaxy, see next section \ref{observations}. 

\section{Astrophysical Observations}\label{observations}

\begin{figure}[t]
\includegraphics[width = 0.5\textwidth]{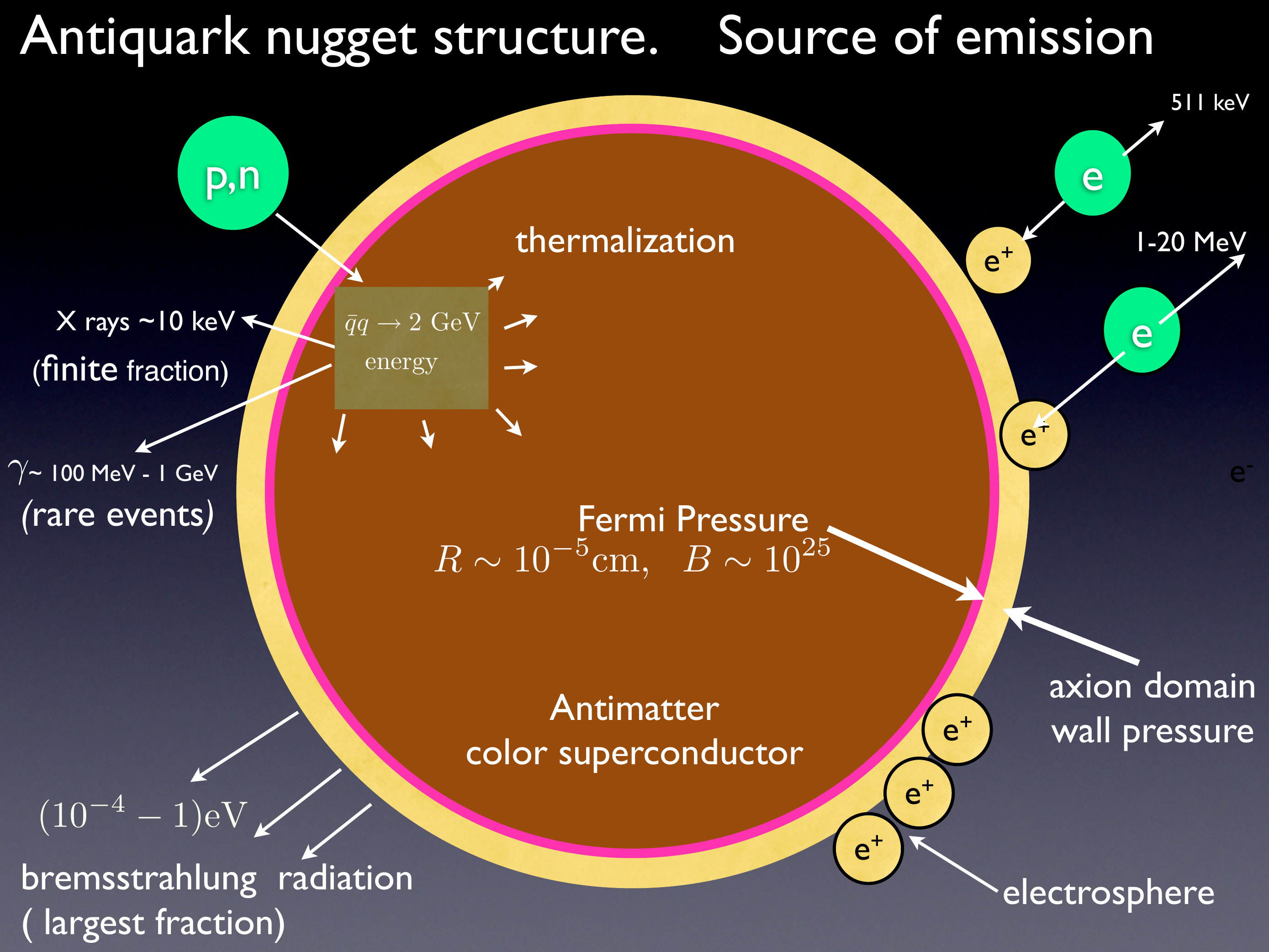}
\caption{ (Anti)nugget internal structure and sources of emission from the core of the galaxy.}
\label{fig:nugget}
\end{figure}

A comparison between different observations of emission 
from the centre  of galaxy is possible because the rate of 
annihilation events is proportional to
$n_{\rm visible}(r)n_{DM}(r)$-the product of the local visible and
 DM distributions at the annihilation site. 
The observed fluxes for different emissions thus depend 
on one and the same line-of-sight integral 
\be
\label{flux}
\Phi \varpropto R^2\int d\Omega dl [n_{\rm visible}(l)\cdot n_{DM}(l)],
\ee
where $R\sim B^{1/3}$ is a typical size of the nugget which determines the 
effective cross section of interaction between DM and visible matter. As 
$n_{DM}\sim B^{-1}$ the effective interaction is strongly suppressed $\sim B^{-1/3}$ 
as we already mention in section \ref{DM}. The average baryonic charge  $B$ 
of the nuggets is the only unknown parameter of the model. It is determined 
by the properties of the axion as reviewed in section \ref{nuggets}.  
In what follows we fix $\Phi$ from (\ref{flux}) for all galactic emissions 
considered below by assuming that our mechanism saturates 511 keV line 
as discussed below. It corresponds to an average baryon charge  $B\sim 10^{25}$ 
for a typical  density distributions  $n_{\rm visible}(r),  n_{DM}(r)$ entering (\ref{flux}). 
Other emissions from different bands  are expressed in terms of the same integral 
(\ref{flux}), and therefore, the  relatives intensities  are completely determined by 
the internal structure of the nuggets which is described by conventional 
nuclear physics and basic QED. 
 
We emphasize that this proposal makes a very  nontrivial prediction:  
the morphology of  all the different  diffuse emission sources discussed 
below must be very strongly correlated.  Furthermore,  in this framework all  
emissions  from different bands are  proportional to 
$\int d l [n_{\rm visible}(l)\cdot n_{DM}(l)]$, which  should be contrasted with many  
other DM models in which the intensities of predicted emissions are proportional 
to $\int d l  n_{DM}^2(l)$ for annihilating DM models or 
$\int d l   n_{DM}(l)$ for decaying DM models.

\begin{figure}[t]
\begin{center}
\includegraphics[width = 0.5\textwidth]{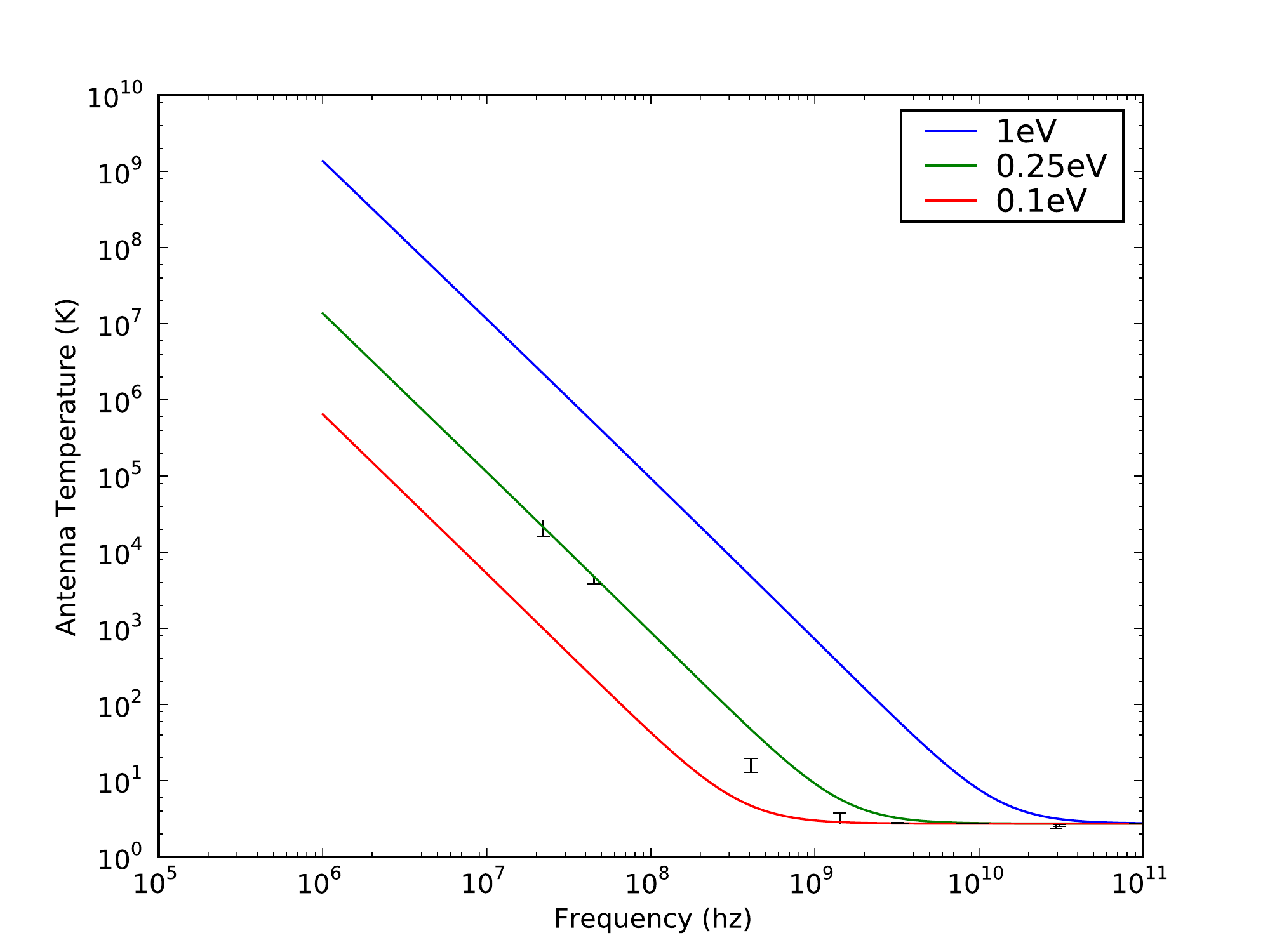}
\caption{Predicted antenna temperature assuming that the quark 
nuggets have a temperature of 0.1eV, 0.25eV and 1eV at the time of \textsc{cmb} 
formation. Also plotted are the data points from the radio band observations.  
Taken from  \cite{Lawson:2012zu}}
\label{ant_temp}
\end{center}
\end{figure}

There are  a number of frequency bands in which an excess  of emission,  
not easily explained by conventional astrophysical sources has been observed.  
These include:
 
a) The \textsc{Spi/Integral} observatory detects a 
stronger than expected 511~keV line associated with the 
galactic centre \cite{Prantzos:2010wi}. 

b) The \textsc{Comptel} satellite observes an excess 
in 1-30~MeV $\gamma$-rays \cite{Strong:2004de}, 
see green vertical bars on Fig.\ref{spectrum}.

c) In the x-ray range \textsc{Chandra} observed 
a $\sim 10$ keV plasma associated with the galactic 
centre. This plasma has no clear heating mechanism and 
is too energetic to remain bound to the galaxy \cite{Muno:2004bs}. 

d) In the microwave range \textsc{WMAP} observes a 
``haze" associated with the foreground galaxy
\cite{Finkbeiner:2003im}. 

e) At temperatures below the CMB peak  
\textsc{Arcade2} has measured a sharp rise in the 
isotropic radio background suggesting an additional 
source of radiation present in the universe before the 
formation of large scale structure \cite{Fixsen:2009xn}, see data points on Fig.\ref{ant_temp}.

The interaction between the nuggets and their environment is governed 
by well known nuclear physics and basic QED. As such their 
observable properties contain relatively few tunable parameters allowing several 
strong tests of the model to be made based on galactic observations. It is found that 
the presence of quark nugget dark matter is not merely allowed by present 
observations but that the overall fit to the diffuse galactic emission  spectrum 
across many orders of magnitude in energy may be improved by their inclusion. 
To be more specific, in our proposal the excesses of  emissions    are explained as follows:   
\begin{figure}[t]
\begin{center}
\includegraphics[width = 0.6\textwidth ]{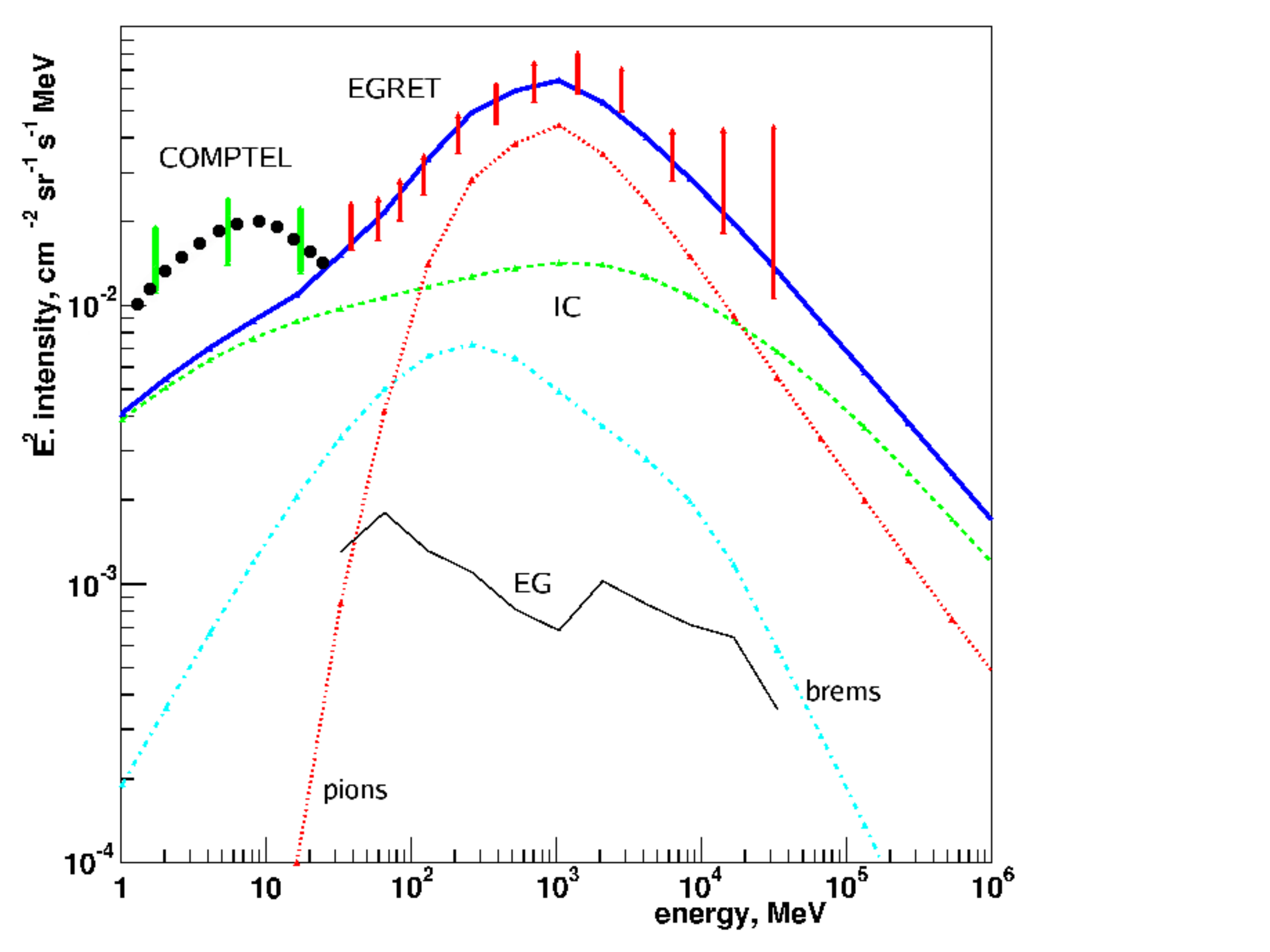}
\caption{$\gamma$ ray spectrum of inner galaxy.   Green vertical bars: COMPTEL data.  Solid blue line: expected total emission due to a combination of conventional astrophysical sources. Heavy black dots: calculated emission spectrum from electron-nugget annihilation processes, taken  from  \cite{Lawson:2007kp}. }
\label{spectrum}
\end{center}
\end{figure}

a) The galactic electrons incident on an antiquark nugget annihilate 
with the surrounding positron layer 
through resonance positronium (Ps) formation.   This 
results in the 511 keV line  with a typical width of order $\sim {\rm few~ keV}$ 
accompanied by the conventional continuum due to $3\gamma$ decay  
\cite{Oaknin:2004mn, Zhitnitsky:2006tu}, 
sketched on  right upper corner on Fig.\ref{fig:nugget}. 
The distribution $[n_{\rm visible}\cdot n_{DM}]$  from eq. (\ref{flux}) implies that the 
predicted emission will be asymmetric, with extension into the disk from the 
galactic center as it tracks the visible matter. There appears to be evidence 
for an asymmetry of this form \cite{Prantzos:2010wi}. 

b)  Some  galactic electrons are able to penetrate to a sufficiently 
large depth  as shown  on right upper corner on Fig.\ref{fig:nugget}. Positrons closer to the quark matter surface can carry energies up 
to the nuclear scale.  These events
 no longer produce  the characteristic positronium decay 
spectrum but a direct non-resonance $e^-e^+ \rightarrow 2\gamma$ emission spectrum 
\cite{Lawson:2007kp}. The transition between these 
two regimes is determined by conventional physics and allows us to  compute   
the strength and spectrum of the MeV scale emissions relative to 
that of the 511~keV line \cite{Forbes:2009wg}. Observations 
by the \textsc{Comptel} satellite indeed show an excess above the galactic 
background \cite{Strong:2004de}  consistent with our estimates, see  heavy black dots on Fig. \ref{spectrum}.
We emphasize that the ratio between these two emissions is determined by well established physics. This ratio is highly sensitive to the    positron density in electrosphere (shown  on right lower corner on Fig.\ref{fig:nugget}) which has highly nontrivial behaviour. It was    was computed using Thomas-Fermi approximation\cite{Forbes:2009wg}  to estimate the spectrum in MeV band shown on Fig.\ref{spectrum}. 

c) Galactic protons incident on the nugget 
will penetrate some distance into the quark matter before 
annihilating into hadronic jets sketched on left upper corner on Fig.\ref{fig:nugget}. This process results in the emission of Bremsstrahlung 
photons at x-ray energies \cite{Forbes:2006ba}. Observations by the 
\textsc{Chandra} observatory indeed indicate an excess in x-ray emissions from 
the galactic centre  \cite{Muno:2004bs} with the intensity and spectrum 
consistent with our estimates \cite{Forbes:2006ba}.

d) Hadronic jets produced 
deeper in the nugget or emitted in the downward direction 
will be completely absorbed. They  eventually emit 
thermal photons with radio frequencies contributing to 
the \textsc{wmap} haze sketched on left lower corner on Fig.\ref{fig:nugget}. Again the relative scales of these 
 emissions may be estimated and is found to be in 
agreement with their observed levels, see
\cite{Forbes:2008uf} for the details.

e)
The source of the emission discussed above with  radio frequencies contributing to 
the WMAP  haze sketched on left lower corner on Fig.\ref{fig:nugget} is also 
 quite active  at earlier times    at $z\sim 10^3$ when the  densities of the particles 
 are about the same order of magnitude as in the center of galaxy at present time. 
 The  analysis \cite{Lawson:2012zu}
finds that at energies near the CMB  peak the nugget contribution to the radio 
background is several orders of magnitude below that of the thermal CMB  spectrum. 
However the CMB  spectrum falls of at frequencies below peak much faster than that 
of the nuggets such that, at frequencies below roughly a Ghz, they come to dominant 
the isotropic radio background. As such the presence of dark matter in the form of 
quark nuggets offers a potential explanation of the radio excess observed by ARCADE2, 
as shown on Fig. \ref{ant_temp}.

These apparent excess emission sources have been cited 
as possible support for a number of dark matter models 
as well as other exotic astrophysical phenomenon. At present 
however they remain open matters for investigation and, given 
the uncertainties in the galactic spectrum and the wide 
variety of proposed explanations are unlikely to provide 
clear evidence in the near future. Therefore, we turn to direct detection prospects of such objects.

\section{Direct Detection Prospects}\label{direct}
Given the uncertainties associated with galactic backgrounds 
a complementary direct detection approach is necessary. 
While direct searches for weakly interacting dark matter 
require large sensitivity a search for high mass dark 
matter requires large area detectors. If the dark matter 
consists of quark nuggets at the $B\sim 10^{25}$ scale they 
will have a flux of
\begin{equation}
\label{eq:flux}
\frac{dN}{dA ~ dt} = nv \approx \left( \frac{10^{25}}{B} \right) {\rm km}^{-2} {\rm yr}^{-1}
\end{equation}
While 
this flux is far below the sensitivity of conventional dark 
matter searches it is similar to the flux of cosmic rays 
near the GZK limit. As such present and future 
experiments investigating ultrahigh energy cosmic rays 
may also serve as search platforms for dark matter of this type.

A nugget of dark matter impacting the earth's  
atmosphere will annihilate the line of atmospheric 
molecules in it's path heating the nugget and depositing 
energy in the atmosphere. For a nugget with a radius of 
$10^{-5}cm$ the annihilation of all molecules along it's
path would result in the annihilation of $10^{-10}kg$ of 
matter and generate $\sim 10^{7}J$ substantially more than 
a conventional UHECR though much of this energy is 
actually thermalized within the nugget. The majority of this energy 
is produced by nuclear annihilations occurring within the  
nugget, the hadronic components released in these events 
remain bound to the quark matter and thermalize 
within it before they are able to escape into the atmosphere. 
While they are less strongly bound electrons are unable 
to escape through the dense positron layer at the nugget 
surface and are also thermalized. As such the emission 
from the nuggets is dominated by relativistic muons and 
thermal photons.  

Recent work has considered the possibility that large scale 
cosmic ray detectors 
may be capable of observing quark nuggets passing through the earth's
atmosphere either through the extensive air shower such an event 
would trigger \cite{Lawson:2010uz} or through the geosynchrotron 
emission generated by the large number of secondary particles
\cite{Lawson:2012vk}. It has also been suggested that the \textsc{anita} 
experiment may be sensitive to the radio band 
thermal emission generated by these objects as they pass through the 
antarctic ice \cite{Gorham:2012hy}. These experiments may thus be 
capable of adding direct detection capability to the indirect evidence 
discussed above in section \ref{observations}.

On entering the earth's crust the nugget will continue to deposit energy along 
its path, however this energy is dissipated in the  surrounding rock and is unlikely 
to be directly observable. Generally the nuggets carry sufficient momentum to travel 
directly through the earth and emerge from the opposite side however a small fraction 
may be captured and deposit all their energy. In  \cite{Gorham:2012hy} the 
possible contribution of energy deposited by quark nuggets to the earth's 
thermal budget was estimated and found to be consistent with observations.

The muonic component of the shower is particularly important 
as it drives an extensive air shower surrounding the quark 
nugget. This shower will be similar to those initiated by 
an ultrahigh energy proton or nucleus, as both arise 
from a large number of hadronic cascades, but with some important 
distinctions.
The most important distinction arises from the 
fact that the quark nugget remains intact as it traverses the 
atmosphere and continues to produce new secondary particles 
all the way to the surface. This introduces a fundamentally new timescale 
into the air shower as the nuggets move significantly slower 
than the speed of light. For nuggets with typical galactic scale 
velocities this implies a slower duration on the 
millisecond scale a thousand times longer than that of a standard 
cosmic ray event. 

As with air showers initiated by a single 
ultra high energy primary these showers may be detected through 
atmospheric fluorescence, surface particle detectors or 
the emission of geosynchrotron radiation in the radio band. 
For a more extensive discussion of the phenomenology 
of quark matter induced shower see \cite{Lawson:2010uz}, 
\cite{Lawson:2012vk}. Both the Pierre Auger Observatory and 
Telescope Array should, in principle, be capable of placing significant 
constraints on the flux of quark nuggets as will the JEM-EUSO 
experiment when it begins taking data. In all cases sensitivity to these 
events will require the analysis of data for events with millisecond 
scale durations. 

\begin{figure}[t]
\begin{center}
\includegraphics[width = 0.5\textwidth]{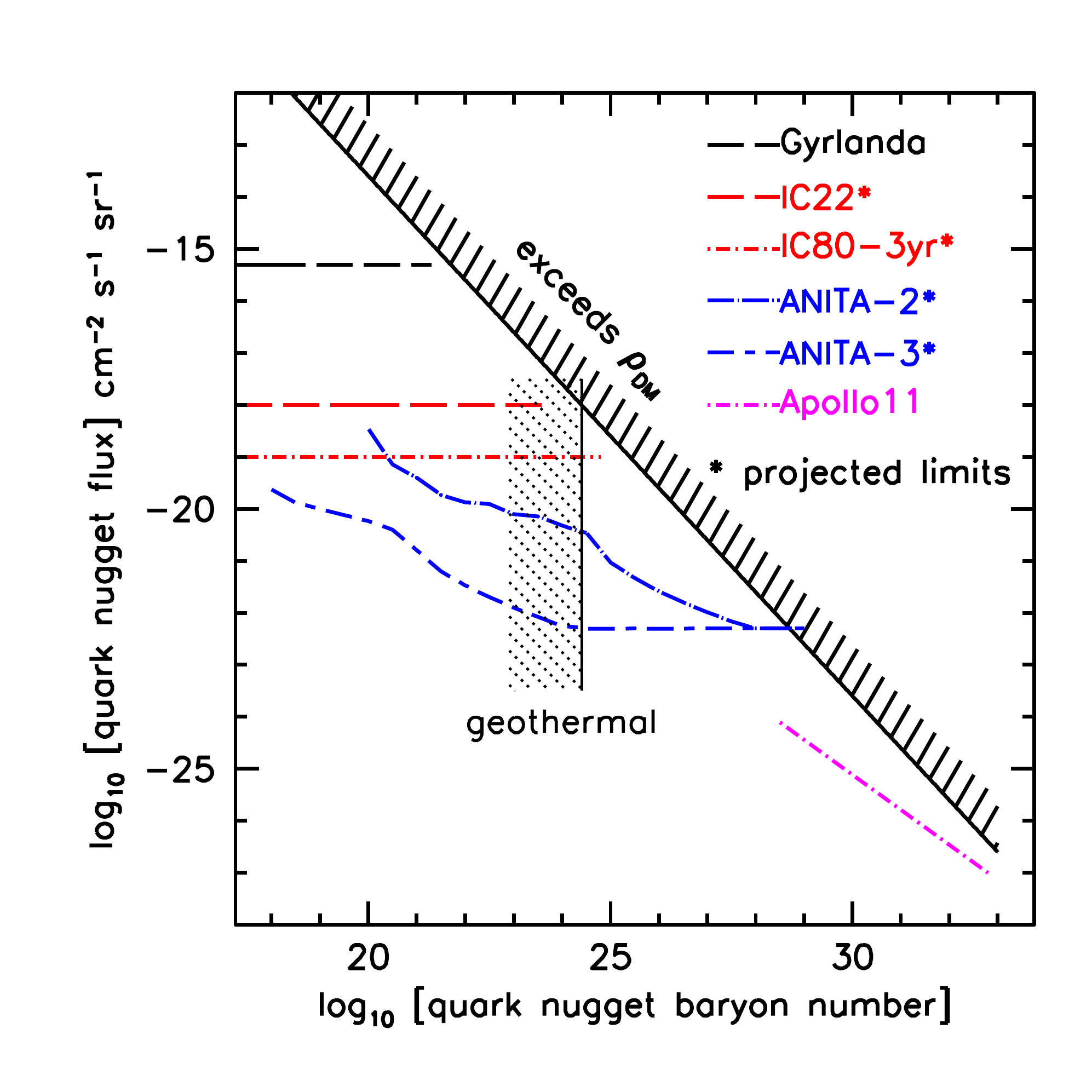}
\caption{ Limits on quark nugget mass and density 
based on current constraints and ANITA data currently 
under analysis. Taken from \cite{Gorham:2012hy}.}
\label{fig:limits}
\end{center}
\end{figure}

A second important differentiating feature of the air shower 
innitiated by a quark nugget is the fact that it will contain 
a thermal component emitted from the nugget surface. 
This temperature will rise with the surrounding density and 
be emitted uniformly in all directions. This distinctly contrasts 
with the highly beamed Cherenkov and geosynchrotron 
radiation associated with traditional cosmic ray showers and 
extending over a much wider frequency range than atmospheric 
fluorescence. At large surrounding densities the nugget temperature can 
easily reach the keV scale and contribute significantly to 
the total emission. In particular the signal  
from a nugget moving through the radio transparrent 
antarctic ice will be dominantly thermal. 
The \textsc{Anita} experiment is sensitive to this radio component and 
the analysis of presently collected data should be able to 
constrain the presence of quark nugget dark matter 
across a significant section of parameter space \cite{Gorham:2012hy} as shown in Fig.\ref{fig:limits}. 

\section{Conclusion}\label{conclusion}
The model which is advocated in the present review was originally invented as a 
simple and natural explanation of  the observed relation: 
$\Omega_{\rm dark} \simeq 5\cdot \Omega_{\rm visible}$ 
by postulating that both elements originated from one and the same QCD scale. 
The immediate consequence of this proposal is the presence   of antimatter 
in a form of macroscopically large  anti-nuggets. 
An equal portion of matter and antimatter in  our universe does not contradict the 
conventional and naive arguments on the near absence of antimatter observed in our
universe as explained in section \ref{DM}. 

It turns out that this  dark matter proposal as a byproduct  
of baryogenesis may explain  a number of  
apparently unrelated puzzles relating to diffuse emission observed in 
many bands as reviewed in section \ref{observations}. 
All these puzzles  strongly suggest (independently)  the presence of some source of
excess diffuse radiation from the centre of galaxy in bands ranging 
over 11 orders of magnitude in frequency. Furthermore,  the 
same dark matter model  can also explain the isotropic background   
observed by \textsc{arcade2} in the radio bands. In this case the emission originates 
primarily from very early times with $z\sim 10^3$ in contrast with our previous 
applications to the 
excess in galactic emissions.  It should be emphasized that all relative 
intensities of diffuse radiation in this proposal 
are fixed as they are determined by conventional physics. The absolute normalization 
is expressed in terms of  a single unknown parameter: the average size of the nugget, 
or identically, the average baryon charge $B\sim 10^{25}$, which itself is 
determined by the axion parameters as explained in section \ref{nuggets}. 

Observation of any morphological correlations between the different excesses 
in diffuse emission mentioned in section \ref{observations} would strongly 
support this proposal as it is difficult to imagine how such a correlation may 
emerge in any other model (which are typically  designed to explain an 
excess of emission in a  single specific frequency  band).

In addition to this type of indirect observational support this model is also amenable to 
direct observational tests. As outlined in section \ref{direct} several large scale 
experiments both, active and planned, have the ability to observe the small but 
non-zero flux of antimatter through the earth. In particular large scale cosmic ray 
detectors intended to study the cosmic rays near the GZK scale are currently 
probing a flux scale comparable to that of quark antinuggets. The passage of 
an antinugget through the atmosphere will produce both electromagnetic radiation 
and a secondary  particle shower that should be observable to a range of cosmic 
ray detectors. The detection and identification of these event will require the 
analysis of data over the millisecond scale typical of a nugget crossing the 
atmosphere (or other large targets such as the antarctic ice).

Finally, what is perhaps more remarkable is the fact that the key assumption of this dark matter model,  the charge separation effect reviewed in sections \ref{DM} and \ref{nuggets},  
can be experimentally tested in heavy ion collisions, where a similar 
$\cal{CP}$ odd environment with $\theta\sim 1$ can be achieved, see  
section IV in ref.\cite{Kharzeev:2007tn} for the details. 
In particular, the local  violation  of the $\cal{CP}$ invariance  observed at RHIC (Relativistic Heavy Ion Collider)\cite{Abelev:2009tx} and LHC (Large Hadron Collider)\cite{Abelev:2012pa}  have been interpreted in  \cite{Kharzeev:2007tn,Zhitnitsky:2010zx,Zhitnitsky:2012im} as an outcome of  a charge separation mechanism in the presence of the induced $\theta\sim 1$   resulting from  a collision. The difference is of course that
 $\cal{CP}$ odd  term with  $\theta\sim 1$  discussed in cosmology describes a theory  on the horizon scale, while $\theta\sim 1$ in heavy ion collisions is correlated on a size of the colliding nuclei. 

 \section*{Acknowledgements}
 We are thankful to participants of the Cosmic Frontier Workshop (CF3 and CF6 groups), SLAC, March 2013,  at which this proposal  was presented,  
for their numerous questions and useful discussions.  
This research was supported in part by the Natural Sciences and 
Engineering Research Council of Canada.


\begin{thebibliography}{10}

\bibitem{Zhitnitsky:2002qa} 
  A.~R.~Zhitnitsky,
  JCAP {\bf 0310}, 010 (2003)
  [hep-ph/0202161].
  
\bibitem{Oaknin:2003uv} 
  D.~H.~Oaknin and A.~Zhitnitsky,
  Phys.\ Rev.\ D {\bf 71}, 023519 (2005)
  [hep-ph/0309086].
  
\bibitem{Zhitnitsky:2006vt} 
  A.~Zhitnitsky,
  Phys.\ Rev.\ D {\bf 74}, 043515 (2006)
  [astro-ph/0603064].
  
\bibitem{Witten:1984rs}
E.~Witten,
\newblock Phys.~Rev.~D {\bf 30}, 272 (1984).

\bibitem{Labun:2011wn} 
  L.~Labun, J.~Birrell and J.~Rafelski,
  Phys.\  Rev.\  Lett.\  110, {\bf 111102} (2013)
  [arXiv:1104.4572 [astro-ph.EP]].


\bibitem{Asztalos:2006kz}
  S.~J.~Asztalos, L.~J.~Rosenberg, K.~van Bibber, P.~Sikivie, K.~Zioutas,
  Ann.\ Rev.\ Nucl.\ Part.\ Sci.\  {\bf 56}, 293-326 (2006).
  
\bibitem{Sikivie:2009fv} 
  P.~Sikivie,
  Int.\ J.\ Mod.\ Phys.\ A {\bf 25}, 554 (2010)
  [arXiv:0909.0949 [hep-ph]].


  
\bibitem{Prantzos:2010wi} 
  N.~Prantzos, C.~Boehm, A.~M.~Bykov, R.~Diehl, K.~Ferriere, N.~Guessoum, P.~Jean and J.~Knoedlseder {\it et al.},
  Rev. \ Mod.\ Phys. {\bf 83},  1001 (2011), 
  arXiv:1009.4620 [astro-ph.HE].

\bibitem{Strong:2004de}
A.~W. Strong, I.~V. Moskalenko, and O.~Reimer,
\newblock Astrophys. J. {\bf 613}, 962 (2004), arXiv:astro-ph/0406254.

\bibitem{Muno:2004bs}
M.~P. Muno {\em et~al.},
\newblock Astrophys. J. {\bf 613}, 326 (2004), arXiv:astro-ph/0402087.

\bibitem{Finkbeiner:2003im}
D.~P. Finkbeiner,
\newblock Astrophys. J. {\bf 614}, 186 (2004), arXiv:astro-ph/0311547.

\bibitem{Fixsen:2009xn}
D.~Fixsen {\em et~al.},
\newblock The Astrophysical Journal {\bf 734} (2011).

\bibitem{Oaknin:2004mn}
D.~H. Oaknin and A.~R. Zhitnitsky,
\newblock Phys.~Rev.~Lett. {\bf 94}, 101301 (2005), arXiv:hep-ph/0406146.

\bibitem{Zhitnitsky:2006tu}
A.~Zhitnitsky,
\newblock Phys.~Rev.~D {\bf 76}, 103518 (2007), arXiv:astro-ph/0607361.

\bibitem{Lawson:2007kp}
K.~Lawson and A.~R. Zhitnitsky,
\newblock JCAP {\bf 0801}, 022 (2008), arXiv:0704.3064 [astro-ph].

\bibitem{Forbes:2009wg} 
  M.~M.~Forbes, K.~Lawson and A.~R.~Zhitnitsky,
  Phys.\ Rev.\ D {\bf 82}, 083510 (2010)
  [arXiv:0910.4541 [astro-ph.GA]].

\bibitem{Forbes:2006ba} 
  M.~M.~Forbes and A.~R.~Zhitnitsky,
  JCAP {\bf 0801}, 023 (2008)
  [astro-ph/0611506].
  
\bibitem{Forbes:2008uf} 
  M.~M.~Forbes and A.~R.~Zhitnitsky,
  Phys.\ Rev.\ D {\bf 78}, 083505 (2008)
  [arXiv:0802.3830 [astro-ph]].
  

\bibitem{Lawson:2012zu} 
  K.~Lawson and A.~R.~Zhitnitsky,
  Phys. Lett. B {\bf 724}, pp. 17 (2013)
  [arXiv:1210.2400 [astro-ph.CO]].
  
 
\bibitem{Lawson:2010uz}
K.~Lawson,
\newblock Phys. Rev. D {\bf 83}, 103520 (2011).

\bibitem{Lawson:2012vk} 
  K.~Lawson,
  Phys.\ Rev.\ D {\bf 88}, 043519 (2013)
  [arXiv:1208.0042 [astro-ph.HE].
  
\bibitem{Gorham:2012hy}
Gorham, Peter.W.,
\newblock Phys. Rev. D {\bf 86}, 123005 (2012).

\bibitem{Kharzeev:2007tn} 
  D.~Kharzeev and A.~Zhitnitsky,
  Nucl.\ Phys.\ A {\bf 797}, 67 (2007)
  [arXiv:0706.1026 [hep-ph]].

\bibitem{Abelev:2009tx}
  B.~I.~Abelev {\it et al.}  [STAR Collaboration],
  Phys.\ Rev.\  C {\bf 81}, 054908 (2010)
  [arXiv:0909.1717 [nucl-ex]].

\bibitem{Abelev:2012pa} 
  B.~Abelev {\it et al.}  [ALICE Collaboration],
  Phys.\ Rev.\ Lett.\  {\bf 110}, 012301 (2013)
  [arXiv:1207.0900 [nucl-ex]].
  
\bibitem{Zhitnitsky:2010zx} 
  A.~R.~Zhitnitsky,
  Nucl.\ Phys.\ A {\bf 853}, 135 (2011)
  [arXiv:1008.3598 [nucl-th]].
  
\bibitem{Zhitnitsky:2012im} 
  A.~R.~Zhitnitsky,
  Nucl.\ Phys.\ A {\bf 886}, 17 (2012)
  [arXiv:1201.2665 [hep-ph]].

\end{thebibliography}
\end{document}